\documentclass
[11pt,tightenlines,eqsecnum,floats,aps,amsmath,amssymb,nofootinbib,prd,shownopacs,
notitlepage]{revtex4}
\usepackage{hyperref}
\usepackage{graphicx}
\usepackage{epstopdf}
\usepackage{amsmath,amssymb,amsfonts,amsthm,latexsym,stmaryrd,mathtools}
\usepackage{dsfont}
\usepackage{pstool}
\def\nn{\nonumber}
\def\be{\begin{equation}}
\def\ee{\end{equation}}
\def\ba{\begin{eqnarray}}
\def\ea{\end{eqnarray}}
\allowdisplaybreaks[4]

\begin{document}

\title{Loop quantum cosmological dynamics of scalar-tensor theory in the Jordan frame}

\author{Yu Han}
 \email{hanyu@xynu.edu.cn}
 \affiliation{College of Physics and Electrical Engineering, Xinyang Normal University, 464000 Xinyang, China}
\date{\today}

\begin{abstract}	
The effective dynamics of scalar-tensor theory (STT) in the Jordan frame is studied in the context of loop quantum cosmology with holonomy corrections. After deriving the effective Hamiltonian from the connection dynamics formulation, we obtain the holonomy-corrected evolution equations of STT on spatially flat Friedmann-Robterson-Walker background, which exhibit some interesting features unique to the Jordan frame of STT. In particular, the linear term of the cosine function appearing in the equations could lead to dynamics much different from the classical theory in the low-energy limit. In the latter part of this paper, we choose a particular model in STT -- the Brans-Dicke theory to specifically illustrate these features. It is found that in Brans-Dicke theory the effective evolution equations can be classified into four different cases. Exact solutions of the Friedmann equation in terms of the internal time are obtained in these cases. Moreover, the solutions in terms of the proper time describing the late time evolution of the Universe are also obtained under certain approximation; in two cases the solutions coincide with the existing solutions in classical Brans-Dicke theory while in the other two cases the solutions do not.
\end{abstract}

\maketitle


\section{Introduction}
The scalar-tensor theory (STT) has been widely investigated in much of the literature during the past several decades, especially in the research with regard to cosmic acceleration in the very early or late Universe (see, for instance,  Refs. \cite{Bergmann:1968,Steinhardt:1990,Garc¨ªa-Bellido:1990,Boisseau:2000,Elizalde:2004,Faraoni:2004}).  In particular, recent astrophysical observations tend to support some inflationary models in STT, which triggers considerable research interest about the effects produced during inflation in STT  (see, for instance, Refs. \cite{Ade:2014,Sebastiani:2014,Aref¡¯eva:2014,Dom¨¨nech:2015}). Nevertheless, the preinflationary evolution in STT which may also leave footprints in observations has largely been neglected yet. In the preinflationary period, due to the extremely high energy and large spacetime curvature, the description by general relativity  may well lose its effectiveness; and to search for the footprints generated during this period, we must fall back on the theory of quantum gravity.

As one of the tentative quantum cosmology theories, loop quantum cosmology (LQC) is  often used to search for the quantum gravity effects \cite{Ashtekar:2006,Ashtekar:2011,Bojowald:2005}. Among the three main quantum corrections in LQC, namely the holonomy correction, inverse-volume correction and quantum backreaction, the holonomy correction, which arises from replacing the classical connection variable by its holonomy around a given square, is usually believed to be dominant when the energy density is much higher than the typical energy scale of slow-roll inflation \cite{Ashtekar:2011,Barrau:2014}. The holonomy correction in STT can be studied either in the Einstein frame or in the Jordan frame.  In the Einstein frame, one first performs the conformal transformation to connection variable and then applies the polymer quantization of LQC to the transformed connection  \cite{Veraguth:2017}, while in the Jordan frame one directly quantizes the connection variable. For simplicity, the effects of holonomy correction in STT were mainly investigated in the Einstein frame in many particular models of STT over the past decade, such as the model with nonminimal coupling
$\xi \phi^2$\cite{Bojowald:2006,Artymowski:2012}, $f(R)$ gravity and Brans-Dicke (BD) theory \cite{Amor¨®s:2014,Odintsov:2014,Jin:2018,Haro:2018}.
In the past few years, the loop quantization of STT has also been formulated in the Jordan frame \cite{Zhang:2011a,Zhang:2011b}; and its cosmological application to BD theory  has been studied in Ref. \cite{Zhang:2013}. The comparison of holonomy correction in  the Einstein frame and the Jordan frame of BD theory from the perspective of effective dynamics was given in Ref. \cite{Artymowski:2013}, in which the author showed that in LQC of BD theory the two frames are no longer equivalent, because the implementation of holonomy correction does not commute with the conformal transformation. Hence, unlike the classical case, the equations in the two frames cannot be switched to each other by conformal transformation. Therefore, the physics in the two frames are completely different, a concrete example of which is that the critical energy density of the scalar fields in BD theory becomes frame dependent. Another crucial difference between the two frames was analyzed for the effective dynamics of $f(R)$ gravity in Ref.
\cite{Chen:2018}, in which $f(R)$ gravity is regarded as a special sector of BD theory; and the author showed that the bounce does not exist in the Jordan frame of $R^2$ gravity for a broad class of initial conditions, which is also completely different from the physics in the Einstein frame in which a bounce generally exists.

Despite these useful results, more work needs to be done in the research of LQC of STT in the Jordan frame. For one thing, only effective dynamics of certain particular models are studied; but the analysis of the general STT is still lacking. For another thing, some existing results in the study of these particular models are still incomplete. To be specific, it is known that in standard LQC with minimally coupled scalar field the function $\cos b$ (or $\sin b$) appears in the effective Hamiltonian in quadratic terms. As a result, only quadratic terms and quartic terms of $\cos b$ (or $\sin b$) are involved in Raychadhuri equation; thus, the sign of $\cos b$ (or $\sin b$) does not directly affect the cosmological evolution. However, in LQC of
BD theory, in addition to the quadratic terms, the Friedmann equation and Raychadhuri equation also involve linear terms of $\cos b$; moreover, as we will show in this paper, the sign of $\cos b$ is to flip around the bounce in BD theory. Thus, whether $\cos b$ takes positive or negative value directly influences the evolution of the Universe. But this crucial fact was not noticed in previous literature such as \cite{Zhang:2013,Artymowski:2013}. To summarize, not only an overall analysis of the effective dynamics with holonomy corrections in the Jordan frame of STT is necessary, but also some existing results should be reanalyzed.

The structure of this paper is as follows. In Sec. II,  the connection dynamics of STT is developed in the Jordan frame. In Sec. III, the corresponding effective Hamiltonian and holonomy-corrected equations of motion in the Jordan frame of STT are derived. In Sec. IV, the BD theory is reanalyzed using the results in Sec. III, exact solutions of effective equations are found. In the last section, we conclude and make some remarks.

\section{Connection dynamics of STT}

 The action of STT in the Jordan frame that we use in this paper reads
\ba
S^{(STT)}=\int_{\Sigma}d^4x\sqrt{|\det(g)|}\bigg[\frac{1}{2\kappa}F(\phi)R
-\frac{1}{2}K(\phi)(\partial^{\mu}\phi)\partial_{\mu}\phi-V(\phi)\bigg],
\label{STTaction}
\ea
 where $\Sigma$ is the spacetime manifold and $\kappa=8\pi G$, and $F(\phi)$, $K(\phi)$  are real-valued, dimensionless functions of the scalar field $\phi$. Besides, in this paper $F(\phi)$, $K(\phi)$  are asked to satisfy
\ba
K(\phi)\neq -\frac{3}{2\kappa} \frac{\big(F'(\phi)\big)^2}{F(\phi)},\label{rq1}
\ea
in which the prime denotes the derivative with respective to $\phi$.
 Otherwise there will be additional constraints in the canonical theory  \cite{Zhang:2011a}, which will much complicate the analysis. Obviously, if $F(\phi)=K(\phi)=1$, the action (\ref{STTaction}) can reproduce the action of general relativity  with a minimally coupled scalar field.

 Note that an alternative form of action of STT often used in the literature is given by
\ba
\tilde{S}^{(STT)}=\frac{1}{2\sqrt{\kappa}}\int_{\Sigma}d^4x\sqrt{|\det(g)|}\bigg[\varphi R-\frac{\omega(\varphi)}{\varphi}(\partial^{\mu}\varphi)\partial_{\mu}\varphi-V(\varphi)\bigg].
\label{action2}
\ea
 At first sight, the action (\ref{STTaction}) can be transformed into the action  (\ref{action2}) through field redefinition $\sqrt{\kappa}\varphi:=F(\phi)$. In fact, the two actions are equivalent to each other only if the function $F(\phi)$ admits an regular inverse $\phi=F^{-1}(\sqrt{\kappa}\varphi)$. However, in many commonly used cosmological models such as the ones in which $F(\phi)$ is expressed by series with even powers of $\phi$, $F(\phi)$ does not have an inverse. In this case,  (\ref{STTaction}) cannot be rewritten in the form of (\ref{action2}). In some literature, the theory of the action (\ref{action2}) is also called ``generalized Brans-Dicke theory." The physical difference between the two actions has been discussed in certain models in the literature such as Refs. \cite{Liddle:1992,Torres:1996}. In this paper, we use action (\ref{STTaction}) as our starting point.

 In the Arnowitt-Deser-Misner (ADM) formulation, the Hamiltonian constraint associated with (\ref{STTaction}) can be expressed in terms of canonical variables as \cite{Han:2015}
 \ba
 \mathcal{H}_{ADM}&=&\frac{1}{\sqrt{\det(q)}}\Bigg[\frac{2\kappa\big(q_{ac}q_{bd}
 -\frac{1}{2}q_{ab}q_{cd}\big)p^{ab}p^{cd}}{F(\phi)}
 +\frac{\big(F'(\phi)q_{ab}p^{cd}-F(\phi)\tilde{\pi}\big)^2}{2F(\phi)G(\phi)}\Bigg]\nn\\
&&+\sqrt{\det(q)}\bigg[-\frac{1}{2\kappa}F(\phi )R^{(3)}+\frac{1}{\kappa}q^{ab}D_a D_b F(\phi )+\frac{K(\phi)}{2}q^{ab}(D_a\phi)D_b\phi+V(\phi)\bigg]\nn\\
&=&0,\label{admconstraint}
 \ea
 where $G(\phi)$ is defined by
 \ba
 G(\phi):=\frac{3}{2\kappa}\big(F'(\phi)\big)^2+F(\phi)K(\phi);\label{Gphi}
 \ea
 and the canonical variables satisfy the standard commutation relationship:
 \ba
 \big\{q_{ab}(x),p^{cd}(y)\big\}=\delta^c_{(a}\delta^d_{b)}\delta^{(3)}(x,y),\quad
  \{\phi(x),\tilde{\pi}(y)\}=\delta^{(3)}(x,y).
 \ea
   In the case $F(\phi)=K(\phi)=1$,  we have $G(\phi)=1$ and (\ref{admconstraint}) reproduces the Hamiltonian constraint in general relativity with a minimally coupled scalar field.

 The ADM phase space can be extended to a larger phase space of connection variables by introducing the su(2)-valued triad $e^a_i$ and its co-triad $e_a^i$ which satisfy
 $q_{ab}=e_a^ie_b^j\delta_{ij}$, $q^{ab}=e^a_ie^b_j\delta^{ij}$. In the new phase space, the densitized triad and its conjugate momentum are defined by
 \ba
 E^a_i:=\sqrt{\det(q)}e^a_i,\qquad K_a^i:=\frac{2\kappa}{\sqrt{\det(q)}}\left(p^{bc}q_{ab}e_c^i
 -\frac{1}{2}\big(p^{bc}q_{bc}\big)e_a^i\right);
 \ea
 and the Ashtekar connection is defined by $A_a^i:=\Gamma_a^i+\gamma K_a^i$, which satisfies \cite{Ashtekar:2004,Thiemann:2007}
 \ba
 \big\{A_a^i(x),E^b_j(y)\big\}=\gamma\kappa\delta_j^i\delta^b_a\delta^{(3)}(x,y),
 \ea
 where $\Gamma_a^i$ is the spin connection compatible with the triad and $\gamma$ is the Barbero-Immirzi parameter.

 The Hamiltonian constraint (\ref{admconstraint}) (modulo the Gauss constraint)  can be reexpressed in terms of the new variables $A,E$ as
 \ba
 \mathcal{H}_{new}&=&\frac{F(\phi)}{2\kappa\sqrt{|\det E}|}E^a_iE^b_j\Bigg[\epsilon^{ij}_{~~k}F^k_{ab}
 -2\bigg(\gamma^2+\frac{1}{(F(\phi))^2}\bigg)K^i_{[a}K^j_{b]}\Bigg]\nn\\
 &&+\frac{1}{2F(\phi)G(\phi)\sqrt{|\det E|}}
 \bigg[\frac{F'(\phi)}{\kappa}(K_a^iE^a_i)
 +F(\phi)\tilde{\pi}\bigg]^2\nn\\
 &&+\sqrt{|\det E|}\bigg[\frac{1}{\kappa}D^a D_a F(\phi )+\frac{K(\phi)}{2}(D^a\phi)D_a\phi+V(\phi)\bigg]\nn\\
 &=&0,\label{newconstraint}
 \ea
 where $F^{~~i}_{ab}:=2\partial_{[a}A^i_{b]}+\epsilon^{~~i}_{jk}A^j_aA^k_b$ is the curvature of Ashtekar connection.
 Note that if we set $F(\phi)=\sqrt{\kappa}\phi$ and $K(\phi)=\frac{\omega(\phi)}{\sqrt{\kappa}\phi}$,  (\ref{newconstraint}) can exactly reproduce the Hamiltonian constraint of generalized Brans-Dicke theory in Ref. \cite{Zhang:2011a}.

  From now on, we consider the spatially flat, homogeneous and isotropic Friedmann-Robertson-Walker (FRW) background. On this background, the line element of the spacetime metric is expressed as
  \ba
  ds^2=-N^2d\tau^2+a^2(dx^2_1+dx^2_2+dx^2_3),\label{FRWmetric}
  \ea
  where $N$ is the homogenous lapse function and $a$ is the scale factor;  and the new variables reduce to
  \ba
  A_a^i=\tilde{c}~{}^o\!e_a^i,\qquad E^a_i=\tilde{p}\sqrt{\det({}^o\!q)}~{}^o\!e^a_i,
  \label{AEFRW}
  \ea
  in which ${}^o\!e^a_i$ and ${}^o\!e_a^i$ represent some fiducial triad and co-triad   and ${}^o\!q_{ab}:={}^o\!e_a^i{}^o\!e_b^j\delta_{ij}$ is the fiducial ADM 3-metric which is related to the physical metric by $q_{ab}=a^2{}^o\!q_{ab}$. Comparing (\ref{AEFRW}) with (\ref{FRWmetric}), we find $|\tilde{p}|=a^2$.

  On the spatially flat FRW background, the Hamiltonian constraint (\ref{newconstraint}) reduces to
  \ba
  \mathcal{H}_{FRW}&=&
  \sqrt{\det({}^o\!q)}
  \bigg[-\frac{3\sqrt{|\tilde{p}|}\tilde{c}^2}{\kappa\gamma^2F(\phi)}
  +\frac{1}{2F(\phi)G(\phi)|\tilde{p}|^{\frac{3}{2}}}
  \Big(\frac{3}{\kappa\gamma}F'(\phi)\tilde{c}\tilde{p}+F(\phi)\tilde{\pi}\Big)^2
  +|\tilde{p}|^{\frac{3}{2}}V(\phi)\bigg]\nn\\
  &=&0.\label{HconstrFRW}
  \ea
  It is convenient of to introduce the following variables which are independent of the fiducial metric:
  \ba
  c:=\mathcal{V}_o^{\frac{1}{3}} \tilde{c},\quad p:=\mathcal{V}_o^{\frac{2}{3}}\tilde{p},
  \quad \pi:=\mathcal{V}_o \tilde{\pi},
  \ea
   where $\mathcal{V}_o:=\int_{\mathcal{C}}d^3x\sqrt{\det({}^o\!q)}$ is the volume of the elementary cell $\mathcal{C}$ measured by the fiducial metric ${}^o\!q_{ab}$.
     With these variables, the background Hamiltonian is given by
   \ba
   H_{FRW}&=&\int_{\mathcal{C}}d^3x N\mathcal{H}_{FRW}\nn\\
  &=&N\Bigg[-\frac{K(\phi)}{G(\phi)}\frac{3\sqrt{|p|}c^2}{\kappa\gamma^2}
  +\frac{F'(\phi)}{G(\phi)}\frac{\text{sgn}(p)}{\sqrt{|p|}}\frac{3c\pi}{\kappa\gamma}
 +\frac{F(\phi)}{G(\phi)}\frac{\pi^2}{2|p|^{\frac{3}{2}}}
  +|p|^{\frac{3}{2}}V(\phi)\Bigg],\label{Hconstraint}
  \ea
  in which the conjugate variables satisfy
  \ba
  \{c,p\}=\frac{\kappa\gamma}{3},\qquad \{\phi,\pi\}=1.
  \ea
 Then, using the Hamilton's equation
 \ba
 \frac{df}{d\tau}=\{f,H_{FRW}\},
 \ea
 we obtain the equations of motion of the canonical variables,
 \ba
 \frac{dc}{d\tau}&=&\frac{N}{2}\frac{\text{sgn}(p)}{\sqrt{|p|}}
 \bigg(-\frac{K(\phi)}{G(\phi)}\frac{c^2}{\gamma}
  -\frac{F'(\phi)}{G(\phi)}\frac{c\pi}{p}
 -\frac{F(\phi)}{G(\phi)}\frac{\kappa\gamma\pi^2}{2p^2}
  +\kappa\gamma |p|V(\phi)\bigg),\label{dc}\\
 \frac{dp}{d\tau}&=&2N\frac{\text{sgn}(p)}{\sqrt{|p|}}
 \bigg(\frac{K(\phi)}{G(\phi)}\frac{cp}{\gamma}
 -\frac{F'(\phi)}{G(\phi)}\frac{\pi}{2}\bigg),\label{dp}\\
 \frac{d\phi}{d\tau}&=&\frac{N}{|p|^{\frac{3}{2}}}
 \bigg(\frac{F'(\phi)}{G(\phi)}\frac{3cp}{\kappa\gamma}+\frac{F(\phi)}{G(\phi)}\pi\bigg),
 \label{dphi}\\
 \frac{d\pi}{d\tau}&=&
 N\sqrt{|p|}\Bigg[\bigg(\frac{K(\phi)}{G(\phi)}\bigg)'\frac{3c^2}{\kappa\gamma^2}
 -\bigg(\frac{F'(\phi)}{G(\phi)}\bigg)'\frac{3c\pi}{\kappa\gamma p}
 -\bigg(\frac{F(\phi)}{G(\phi)}\bigg)'\frac{\pi^2}{2|p|^{2}}
 -|p|V'(\phi)\Bigg],\label{dpi}
 \ea
 in which $\text{sgn}(p)$ is the sign function of $p$, which is related to the scale factor by $|p|=a^2\mathcal{V}^{\frac{2}{3}}_o$. After setting $N=1$, the coordinate time ``$d\tau$" becomes the proper time ``$dt$". In the following, we use ``$\cdot$" to denote the differentiation with respect to the proper time; thus, Eq.
 (\ref{dphi}) becomes
 \ba
 \dot{\phi}&=&\frac{1}{G(\phi)|p|^{\frac{3}{2}}}
 \bigg(\frac{3}{\kappa\gamma}F'(\phi)cp+F(\phi)\pi\bigg).\label{phidot}
 \ea
 Plugging Eq. (\ref{phidot}) into the smeared Hamiltonian constraint (\ref{Hconstraint}), we obtain
 \ba
 \frac{c^2}{\gamma^2|p|}=\frac{G(\phi)}{2}\dot{\phi}^2+F(\phi)V(\phi);\label{constraint2}
 \ea
 then, by using Eq. (\ref{dp}), the constraint (\ref{constraint2}) gives the Friedmann equation of STT,
\ba
F(\phi)H^2+\dot{F}(\phi)H=
\frac{\kappa}{3}\bigg(\frac{K(\phi)}{2}\dot{\phi}^2+V(\phi)\bigg),\label{cFreq}
\ea
 where $H$ is the Hubble parameter,
\ba
H:=\frac{\dot{p}}{2p}.
\ea

Using the canonical equations of motion, it is straightforward to derive the second-order evolution equations. First, from the Friedmann equation (\ref{cFreq}) and the Eqs. (\ref{dc}), (\ref{dphi}),  we derive the Raychadhuri equation
\ba
F(\phi)\dot{H}-\frac{1}{2}\dot{F}(\phi)H=-\frac{1}{2}\Big(\kappa K(\phi)\dot{\phi}^2
+\ddot{F}(\phi)\Big);\label{cRaeq}
\ea
then, by taking time derivative of (\ref{cFreq}) and using (\ref{cRaeq}), we obtain the Klein-Gordon equation,
\ba
\ddot{\phi}+3H\dot{\phi}+\frac{1}{2}\frac{\dot{G}(\phi)}{G(\phi)}\dot{\phi}
-\frac{1}{G(\phi)}\Big[2F'(\phi)V(\phi)-F(\phi)V'(\phi)\Big]=0.\label{cKGeq}
\ea

 Now let us take a further look at the above equations of motion. It is interesting to observe that the theory allows the existence of bounce, at which
\ba
 H=0,\qquad \dot{H}>0.
\ea
From the Eqs. (\ref{cFreq}) and (\ref{cRaeq}), it is easy to see that at the bouncing point the following two conditions have to be satisfied:
\ba
 \frac{K(\phi)}{2}\dot{\phi}^2+V(\phi)=0,\qquad \frac{1}{F(\phi)}\Big(\kappa K(\phi)\dot{\phi}^2+\ddot{F}(\phi)\Big)<0.\label{bouncecon}
\ea
 The conditions in (\ref{bouncecon}) can be met in some models in STT, because from (\ref{constraint2}) we see that the effective kinetic energy term in STT is $\frac{1}{2}G(\phi)\dot{\phi}^2$ and the effective potential is $F(\phi)V(\phi)$, and thus the functions $F(\phi)$, $K(\phi)$ and $V(\phi)$ can be negative to meet the conditions in (\ref{bouncecon}), as long as the right-hand side of the constraint (\ref{constraint2}) remains positive. The bouncing behavior in STT has been of much theoretical interest in the cosmology (see the recent articles \cite{Boisseau:2015,Pozdeeva:2016} for reference). Note that in the minimally coupled case with $F(\phi)=K(\phi)=1$, the bounce can never exist, because the second inequality in (\ref{bouncecon}) is violated.

\section{Effective dynamics of STT with holonomy corrections}
After briefly analyzing the classical dynamics of STT, we proceed to study the quantum gravity effects in STT. Generally speaking, it is necessary to exploit the quantum constraint equation to thoroughly understand the theory on the quantum level. However, this usually requires complicated numerical analysis, especially in our case with nonminimally coupling functions. To avoid the involved numerical analysis and at the same time capture the essential features of quantum corrections, we exploit the effective dynamics to study STT on the semiclassical level (just like what has been done for BD theory in \cite{Zhang:2013}). In the previous part of this section, we promote the Hamiltonian constraint to a quantum operator and derive the effective Hamiltonian using path integral in the timeless framework of LQC; in the latter part, by using the effective Hamiltonian, we obtain the first order canonical equations, from which the second-order evolution equations can be subsequently derived. Some novel features of the holonomy corrections in the Jordan frame are also discussed.

\subsection{Effective Hamiltonian of STT}

To study the quantum dynamics of STT, first we should have well-defined operators in LQC. Note that in the Hamiltonian constraint (\ref{newconstraint}) there are both quadratic and linear terms of the connection. However, in loop quantum gravity there is no direct analog of the connection operator; instead, we only have well-defined holonomy operators. To have a plausible relation to the full theory, in LQC we need to express these terms as functions of holonomies. In this paper, we use the polymerlike quantization prescription first put forward in \cite{Ashtekar:2009} and intensively studied in \cite{Wilson-Ewing:2010,Corichi:2011}, in which the connection operator on the spatially homogenous background is expressed by
\ba
\hat{c}=
\widehat{\frac{\sin (\bar{\mu}c)}{\bar{\mu}}},\label{opA}
\ea
where $\bar{\mu}$ stands for the length of the curve used to calculate the holonomy along it, and $\bar{\mu}=\sqrt{\frac{\Delta}{|p|}}$ with $\Delta=4\sqrt{3}\pi\gamma G\hbar$ being the minimum nonzero eigenvalue of the area operator in loop quantum gravity. For convenience, we introduce the new conjugate variables:
\ba
b:=\bar{\mu}c,\quad v:=2\sqrt{3}~\text{sgn}(p)\bar{\mu}^{-3},
\ea
 where $v$ is proportional to the physical volume of the elementary cell, satisfying $\{b,v\}=\frac{2}{\hbar}$.

For the quantization of the scalar field, we use the standard Schr\"{o}dinger representation adopted in \cite{Ashtekar:2006,Zhang:2013}, in which the kinematical Hilbert space for the scalar field is constructed as in standard quantum mechanics.
Now, the whole kinematical Hilbert space of STT $\mathcal{H}^{STT}_{kin}$ becomes a direct product of the Hilbert space of the geometry and that of the scalar field. We denote the orthonormal basis for $\mathcal{H}^{STT}_{kin}$ by $|v,\phi\rangle$, on which the operator $\hat{v}$ acts by simple multiplication and the operator $\widehat{\sin b}$ acts by \cite{Corichi:2011}
\ba
\widehat
{\sin b}|v,\phi\rangle=\frac{1}{2i}\left[|v+2,\phi\rangle-|v-2,\phi\rangle\right].
\label{actsinb}
\ea
In the Schr\"{o}dinger representation for the scalar field, the operators $\widehat{\left(\frac{K(\phi)}{G(\phi)}\right)}$,
$\widehat{\left(\frac{F'(\phi)}{G(\phi)}\right)}$, $\widehat{\left(\frac{F(\phi)}{G(\phi)}\right)}$,
$\hat{V}(\phi)$ in the quantum Hamiltonian constraint also act on $|v,\phi\rangle$  by multiplication, for instance,
\ba
\widehat{\left(\frac{K(\phi)}{G(\phi)}\right)}|v,\phi\rangle=\frac{K(\phi)}{G(\phi)}|v,\phi\rangle;
\label{action3}
\ea
and the conjugate momentum of $\phi$ acts by differentiation.
 It is obvious that such definition of operators is well defined only if the functions $\frac{K(\phi)}{G(\phi)}$,  $\frac{F'(\phi)}{G(\phi)}$, $\frac{F(\phi)}{G(\phi)}$ and $V(\phi)$ have no singularities for any $\phi$.

Moreover, in order to avoid the quantization ambiguities caused by the inverse-volume operator, following \cite{Ashtekar:2011}, we set $N=|p|^{\frac{3}{2}}$ prior to quantization; then, by using (\ref{opA}), we obtain a symmetric expression of the Hamiltonian constraint operator,
\ba
\hat{H}_{FRW}&=&-\frac{\Delta^2}{4\kappa\gamma^2}\widehat{\left(\frac{K(\phi)}{G(\phi)}\right)}
\hat{v}(\widehat{\sin b})^2\hat{v}
+\frac{3\hbar}{16}
\left[\widehat{\left(\frac{F'(\phi)}{G(\phi)}\right)}\hat{\pi}+\hat{\pi}\widehat{\left(\frac{F'(\phi)}{G(\phi)}\right)}\right]
\left[\widehat{\sin b}\hat{v}+\hat{v}\widehat{\sin b}\right]  \nn\\
&&+\frac{1}{4}\left[\hat{\pi}^2\widehat{\left(\frac{F(\phi)}{G(\phi)}\right)}
+\widehat{\left(\frac{F(\phi)}{G(\phi)}\right)}\hat{\pi}^2\right]
+\frac{(\Delta)^{3}}{12}\hat{v}^2\hat{V}(\phi)\nn\\
&=:&\hat{H}_{1}+\hat{H}_{2}+\hat{H}_{3}+\hat{H}_{4}.\label{QHconstraint}
\ea
 It is worth mentioning that the change of factor ordering in (\ref{QHconstraint}) only affects the details of the theory but does not affect the general properties of effective dynamics. Using (\ref{actsinb}), the action of $\hat{H}_1$ and $\hat{H}_2$ on the quantum state read separately as
\ba
\hat{H}_{1}|v,\phi\rangle&=&
\frac{\Delta^2}{16\kappa\gamma^2}\frac{K(\phi)}{G(\phi)}v\left[(v+4)|v+4,\phi\rangle
-2v|v,\phi\rangle+(v-4)|v-4,\phi\rangle\right],\label{actH1}\\
\hat{H}_{2}|v,\phi\rangle&=&
-\frac{3i\hbar}{16}
\left[\widehat{\left(\frac{F'(\phi)}{G(\phi)}\right)}\hat{\pi}+\hat{\pi}\frac{F'(\phi)}{G(\phi)}\right]
\big[(v+1)|v+2,\phi\rangle-(v-1)|v-2,\phi\rangle\big].\label{actH2}
\ea
Now, let us derive the effective Hamiltonian using the timeless path integral approach developed in \cite{Ashtekar:2010}. In this approach, the transition amplitude in the traditional path integral is replaced by the following extraction amplitude which can extract physical states from kinematical states in the Hilbert space,
\ba
A(v_f,\phi_f;v_i,\phi_i)&:=&\int d\alpha\langle
v_f,\phi_f|e^{\frac{i}{\hbar}\alpha\hat{H}}|v_i,\phi_i\rangle\nn\\
&=&\int d\alpha\sum_{v_{N-1},...,v_1}\int d\phi_{N-1}...d\phi_1
\langle v_N,\phi_N|e^{\frac{i}{\hbar}\epsilon\alpha\hat{H}}|v_{N-1},\phi_{N-1}\rangle...
\langle v_1,\phi_1|e^{\frac{i}{\hbar}\epsilon\alpha\hat{H}}|v_{0},\phi_{0}\rangle,\nn\\
\label{path1}
\ea
 in which we have decomposed the extraction amplitude into $N$ parts with $\epsilon=\frac{1}{N}$, and $\langle v_N,\phi_N|\equiv\langle v_f,\phi_f|$, $|v_{0},\phi_{0}\rangle\equiv|v_{i},\phi_{i}\rangle$. For each part with $\epsilon\ll1$, we have
\ba
\langle v_{n+1},\phi_{n+1}|e^{\frac{i}{\hbar}\epsilon\alpha\hat{H}}|v_{n},\phi_{n}\rangle=
\delta(\phi_{n+1},\phi_{n})\delta_{v_{n+1},v_{n}}
+\frac{i}{\hbar}\epsilon\alpha\sum_{i=1}^{4}\langle v_{n+1},\phi_{n+1}|\hat{H}_i|v_{n},\phi_{n}\rangle
+\mathcal{O}(\epsilon^2).\label{path2}
\ea
 Using Eqs. (\ref{actH1}) and (\ref{actH2}), we can calculate the matrix elements in (\ref{path2}). First, we have
\ba
\langle v_{n+1},\phi_{n+1}|\hat{H}_1|v_{n},\phi_{n}\rangle&=&
\frac{\Delta^2}{16\kappa\gamma^2}\frac{K(\phi_{n})}{G(\phi_{n})}\delta(\phi_{n+1},\phi_{n})
v_nv_{n+1}(\delta_{v_{n+1},v_{n}+4}-2\delta_{v_{n+1},v_{n}}+\delta_{v_{n+1},v_{n}-4})\nn\\
&=&-\frac{\Delta^2}{16\pi^2\kappa\gamma^2\hbar}v_nv_{n+1}\frac{K(\phi_{n})}{G(\phi_{n})}
\int d\pi_{n+1}
\exp\left[\frac{i}{\hbar}\epsilon\pi_{n+1}\frac{\phi_{n+1}-\phi_n}{\epsilon}\right]\nn\\
&&\times\int_{-\pi}^{\pi}d b_{n+1}\sin^2b_{n+1}\exp\left[-\frac{i}{2}\epsilon b_{n+1}\frac{v_{n+1}-v_{n}}{\epsilon}\right],\label{meH1}
\ea
where in the last step we have used the identities:
\ba
\delta_{v,v'}=\frac{1}{2\pi}\int^{\pi}_{-\pi}db\exp[-\frac{i}{2}b
(v-v')],\qquad \delta(\phi,\phi')=
\frac{1}{2\pi\hbar}\int d\pi\exp\left[\frac{i}{\hbar}\pi(\phi-\phi')\right];\label{ids}
\ea
then, by using (\ref{ids}) again, we can calculate the remaining matrix elements in (\ref{path2}), which, together with (\ref{meH1}), yield the result:
\ba
A(v_f,\phi_f;v_i,\phi_i)&=&\int d\alpha\sum_{v_{N-1},...,v_1}
\Big(\frac{1}{2\pi}\Big)^N\int^{\pi}_{-\pi} db_N...db_1
\int d\phi_{N-1}...d\phi_{1}
\Big(\frac{1}{2\pi\hbar}\Big)^N\int d\pi_{N}...d\pi_{1}\nn\\
&&\times\exp(\frac{i}{\hbar}S_N),\label{transam}
\ea
where
\ba
S_N&=&\epsilon \sum_{n=0}^{N-1}\Bigg[\pi_{n+1}\frac{\phi_{n+1}-\phi_n}{\epsilon}
-\frac{\hbar}{2}b_{n+1}\frac{v_{n+1}-v_{n}}{\epsilon}
+\alpha\Bigg(-\frac{\Delta^2}{4\kappa\gamma^2}v_nv_{n+1}\frac{K(\phi_{n})}{G(\phi_{n})}\sin^2b_{n+1}\nn\\
&&\qquad\quad
+\frac{3\hbar}{16}\bigg(\frac{F'(\phi_{n+1})}{G(\phi_{n+1})}
+\frac{F'(\phi_n)}{G(\phi_n)}\bigg)(v_{n+1}+v_n)\pi_{n+1}\sin b_{n+1}\nn\\
&&\qquad\quad
+\frac{1}{4}\bigg(\frac{F(\phi_{n+1})}{G(\phi_{n+1})}+\frac{F(\phi_n)}{G(\phi_n)}\bigg)\pi^2_{n+1}
+\frac{(\Delta)^{3}}{12}v^2_{n+1}V(\phi_{n+1})\Bigg)\Bigg].\label{effaction}
\ea
In the continuum limit with $N\rightarrow \infty$, the extraction amplitude (\ref{transam}) can be expressed as
\ba
A(v_f,\phi_f;v_i,\phi_i)=\int \mathcal{D}\alpha\int \mathcal{D}v\int \mathcal{D}b
\int \mathcal{D}\phi\int \mathcal{D}\pi\exp[\frac{i}{\hbar}\tilde{S}],
\ea
where
\ba
\tilde{S}&=&\int_{0}^{1}d\tau\Bigg[\pi\dot{\phi}-\frac{\hbar}{2}b\dot{v}
+\alpha\Bigg(-\frac{\Delta^2}{4\kappa\gamma^2}\frac{K(\phi)}{G(\phi)}v^2\sin^2b
+\frac{3\hbar}{4}\frac{F'(\phi)}{G(\phi)}\pi v\sin b
+\frac{1}{2}\frac{F(\phi)}{G(\phi)}\pi^2\nn\\
&&\qquad\quad+\frac{(\Delta)^{3}}{12}v^2V(\phi)\Bigg)\Bigg].\label{effaction2}
\ea
From (\ref{effaction2}), it is direct to read off the effective Hamiltonian constraint. Recall that we have set $N=|p|^{\frac{3}{2}}$ for  convenience of quantization. To describe the realistic evolution of the Universe, we reset $N=1$ in the following; hence, the effective Hamiltonian constraint is given by
\ba
H_{eff}&=&-\frac{\sqrt{3\Delta}}{2\kappa\gamma^2}\frac{K(\phi)}{G(\phi)}|v|\sin^2 b
+\text{sgn}(v)\frac{3\sqrt{3}\hbar}{2(\Delta)^{\frac{3}{2}}}
\frac{F'(\phi)}{G(\phi)}\pi\sin b
+\frac{\sqrt{3}}{(\Delta)^{\frac{3}{2}}}\frac{F(\phi)}{G(\phi)}\frac{\pi^2}{|v|}
+\frac{(\Delta)^{\frac{3}{2}}}{2\sqrt{3}}|v|V(\phi)\nn\\
&=&0.
\label{effH}
\ea

  Finally, we mention again that the above derivation of effective Hamiltonian constraint requires that the functions $\frac{K(\phi)}{G(\phi)}$, $\frac{F'(\phi)}{G(\phi)}$, $\frac{F(\phi)}{G(\phi)}$, $V(\phi)$ are continuous and differential functions for all $\phi$. However, in some particular models of STT, this requirement cannot be satisfied. In this case, the path integral will diverge at the singularities of these functions; thus, the form of effective Hamiltonian constraint (\ref{effH}) may not be applicable, and one may need to find alternative definitions of operators to remove the singularities in the path integral.

\subsection{Effective equations of motion}

 Using the effective Hamiltonian $H_{eff}$, the Hamilton's equations of motion of the canonical variables can be directly obtained,
\ba
\dot{b}&=&\text{sgn}(v)\frac{2}{\hbar}
\Bigg[-\frac{\sqrt{3\Delta}}{2\kappa\gamma^2}\frac{K(\phi)}{G(\phi)}\sin^2 b
-\frac{\sqrt{3}}{(\Delta)^{\frac{3}{2}}}\frac{F(\phi)}{G(\phi)}\frac{\pi^2}{|v|^2}
+\frac{(\Delta)^{\frac{3}{2}}}{2\sqrt{3}}V(\phi)\Bigg],\label{eomb}\\
\dot{v}&=&\text{sgn}(v)\frac{2}{\hbar}
\Bigg[\frac{\sqrt{3\Delta}}{\kappa\gamma^2}\frac{K(\phi)}{G(\phi)}v\sin b\cos b
-\frac{3\sqrt{3}\hbar}{2(\Delta)^{\frac{3}{2}}}
\frac{F'(\phi)}{G(\phi)}\pi\cos b \Bigg],\label{eomv}\\
\dot{\phi}&=&\text{sgn}(v)\frac{3\sqrt{3}\hbar}{2(\Delta)^{\frac{3}{2}}}
\frac{F'(\phi)}{G(\phi)}\sin b
+\frac{2\sqrt{3}}{(\Delta)^{\frac{3}{2}}}\frac{F(\phi)}{G(\phi)}\frac{\pi}{|v|},\label{eomphi}\\
\dot{\pi}&=&\frac{\sqrt{3\Delta}}{2\kappa\gamma^2}
\bigg(\frac{K(\phi)}{G(\phi)}\bigg)'|v|\sin^2 b
-\text{sgn}(v)\frac{3\sqrt{3}\hbar}{2(\Delta)^{\frac{3}{2}}}
\bigg(\frac{F'(\phi)}{G(\phi)}\bigg)'\pi\sin b
-\frac{\sqrt{3}}{(\Delta)^{\frac{3}{2}}}\bigg(\frac{F(\phi)}{G(\phi)}\bigg)'\frac{\pi^2}{|v|}
\nn\\
&&-\frac{(\Delta)^{\frac{3}{2}}}{2\sqrt{3}}|v|V'(\phi).\label{eompi}
\ea
These canonical equations can be exploited to derive the Friedmann equation. From (\ref{eomphi}), we have
\ba
\frac{F(\phi)}{G(\phi)}\frac{\pi}{|v|}=\frac{(\Delta)^{\frac{3}{2}}}{2\sqrt{3}}
\bigg(\dot{\phi}-\text{sgn}(v)\frac{3\sqrt{3}\hbar}{2(\Delta)^{\frac{3}{2}}}
\frac{F'(\phi)}{G(\phi)}\sin b\bigg),\label{eqpi}
\ea
substitute (\ref{eqpi}) into (\ref{eomv}), we obtain
\ba
HF(\phi)=\frac{1}{3}\frac{\dot{v}}{v}F(\phi)
=\frac{2\sqrt{3\Delta}}{3\kappa\gamma^2\hbar}\text{sgn}(v)\sin b\cos b
-\frac{1}{2}\dot{F}(\phi)\cos b,\label{HF}
\ea
from which we find that the Hubble parameter will vanish at
\ba
\sin b=\frac{3\kappa\gamma^2\hbar}{4\sqrt{3\Delta}}\text{sgn}(v)\dot{F}(\phi),
\ea
or at
\ba
\cos b=0;
\ea
then, inserting (\ref{eqpi}) into (\ref{effH}), we get
\ba
\sin^2b=\frac{\rho_e}{\rho_c},\label{sin2b}
\ea
where $\rho_c\equiv \frac{3}{\Delta\kappa\gamma^2}$ and the effective energy density $\rho_e$ is defined by
\ba
\rho_e:=\frac{G(\phi)}{2}\dot{\phi}^2+F(\phi)V(\phi).\label{rhoe}
\ea
Equation (\ref{sin2b}) implies that the effective energy density is upper bounded by $\rho_c$; and Eq. (\ref{HF}) implies that the Hubble parameter will vanish if $\rho_e$ reaches $\rho_c$. Combination of Eqs. (\ref{HF}) and (\ref{sin2b}) gives the Friedmann equation in LQC of STT,
\ba
F(\phi)H^2+\dot{F}(\phi)H\cos b=
\frac{\kappa}{3}\bigg(\frac{K(\phi)}{2}\dot{\phi}^2+V(\phi)\bigg)\cos^2b.\label{qFr}
\ea
  Multiplying $F(\phi)$ on both sides of (\ref{qFr}) and inserting
 \ba
 \cos^2b=1-\frac{\rho_e}{\rho_c}, \label{cos2b}
 \ea
 we obtain
\ba
\bigg(F(\phi)H+\frac{1}{2}\dot{F}(\phi)\cos b\bigg)^2
=\frac{\kappa}{3}\rho_e\bigg(1-\frac{\rho_e}{\rho_c}\bigg).\label{qFr2}
\ea
It is easy to check that Eq. (\ref{qFr2}) can reproduce the Friedmann equation (\ref{qFr}) when $F(\phi)\neq0$.
Since $\rho_e$ is bounded from above by $\rho_c$, from Eq. (\ref{qFr2}) we learn that the Hubble parameter cannot approach infinity during the whole evolution.

In the phase space of a collapsing Universe, if the point $\cos b=0$ can be reached, the Universe will stop collapsing; then, if $\dot{b}\neq0$, we have $\dot{\cos b}\neq0$, and thus the sign of $\cos b$ will flip around $\cos b=0$. According to Eq. (\ref{HF}), the Hubble parameter will also change its sign around this point; in other words, a bounce will take place. We call such a bounce the ``quantum bounce" to distinguish from the bounce mentioned in Sec. II which could happen at a much lower energy density than $\rho_c$. Owing to the complicated theoretical structure, in STT a collapsing Universe is not necessarily followed by the quantum bounce.

 Using the effective Hamiltonian constraint (\ref{effH}) and the Hamilton's equations of motion (\ref{eomb})-(\ref{eompi}), after long but straightforward derivations, we obtain the effective Klein-Gordon equation in STT,
 \ba
 \ddot{\phi}+3H\dot{\phi}+\frac{1}{2}\frac{\dot{G}(\phi)}{G(\phi)}\dot{\phi}
-\frac{1}{G(\phi)}\Big[(3\cos b-1)F'(\phi)V(\phi)-F(\phi)V'(\phi)\Big]=0.\label{qKGeq}
\ea
Moreover, by taking time derivative of Eqs. (\ref{qFr}), (\ref{cos2b}) and using Eq. (\ref{qKGeq}), tedious calculations yield the effective Raychadhuri equation in STT,
\ba
F(\phi)\dot{H}\cos b-\dot{F}(\phi)H\bigg(\frac{3}{2}\cos 2b-\cos b\bigg)
=-\frac{1}{2}\Big(\kappa K(\phi)\dot{\phi}^2\cos 2b
+\ddot{F}(\phi)\cos b+\dot{F}(\phi)\dot{(\cos b)}\Big)\cos b.\nn\\\label{qRaeq}
\ea

When $F(\phi)=K(\phi)=1$, it is easy to see that the above effective equations in STT turn into the effective equations in LQC with minimally coupled scalar field.

From (\ref{cos2b}), we find $\cos b=\pm \sqrt{1-\frac{\rho_e}{\rho_c}}$, i.e.  the cosine function can take positive or negative values. In the low-energy limit with $\rho_e\rightarrow0$, we have $\cos b\rightarrow1$ or $\cos b\rightarrow-1$ . In the limit $\cos b\rightarrow1$, we can directly check that the effective Friedmann equation (\ref{qFr}), effective Klein-Gordon equation (\ref{qKGeq}) and effective Raychaudhuri equation (\ref{qRaeq}) can separately reduce to their classical counterparts in (\ref{cFreq}), (\ref{cKGeq}), (\ref{cRaeq}). In the other low-energy limit with $\cos b\rightarrow-1$, the effective Friedmann equation (\ref{qFr}) reduces to
\ba
F(\phi)H^2-\dot{F}(\phi)H=
\frac{\kappa}{3}\bigg(\frac{K(\phi)}{2}\dot{\phi}^2+V(\phi)\bigg),\label{lelFreq}
\ea
which can also be transformed into
\ba
\bigg(F(\phi)H-\frac{1}{2}\dot{F}(\phi)\bigg)^2
=\frac{\kappa}{3}\rho_e;\label{lelqFr2}
\ea
and the effective Klein-Gordon equation (\ref{qKGeq}) and Raychaudhuri equation separately reduce to
\ba
\ddot{\phi}+3H\dot{\phi}+\frac{1}{2}\frac{\dot{G}(\phi)}{G(\phi)}\dot{\phi}
+\frac{1}{G(\phi)}\Big[4F'(\phi)V(\phi)+F(\phi)V'(\phi)\Big]=0,\label{lelKGeq}
\ea
and
\ba
F(\phi)\dot{H}+\frac{5}{2}\dot{F}(\phi)H=-\frac{1}{2}\Big(\kappa K(\phi)\dot{\phi}^2
-\ddot{F}(\phi)\Big).\label{lelRaeq}
\ea

Hence, due to the appearance of linear terms of $\cos b$, there exist two different sets of evolution equations in the low-energy limit,
  which is totally different from the minimally coupled LQC where effective equations involve only quadratic or quartic terms of $\cos b$ (or $\sin b$) and thus can reduce to only one set of evolution equations in the low- energy limit. This is a nontrivial quantum gravity effect caused by holonomy corrections in STT, which deeply reflects the fact that the holonomy plays a fundamental role in LQC.

Since the Eqs. (\ref{lelFreq}), (\ref{lelKGeq}) and (\ref{lelRaeq}) do not correspond to their classical counterparts, very naturally the question arises: Does there exist an effective action from which these equations can be derived? After careful exploration, we obtain a negative answer to this question. Nevertheless, there exists an action which can yield the above equations in the slow-roll limit. Consider the following effective action:
\ba
S^{(eff)}=\int_{\Sigma}d^4x\sqrt{|\det(g)|}\bigg(\frac{1}{F(\phi)}\bigg)^{5}\bigg[\frac{1}{2\kappa}R
-\frac{1}{2}\frac{K(\phi)}{F(\phi)}(\partial^{\mu}\phi)\partial_{\mu}\phi
-\frac{V(\phi)}{F(\phi)}\bigg].
\label{effaction3}
\ea
Under the slow-roll condition with
\ba
\left|\frac{\dot{H}}{H^2}\right|\ll1,\quad \left|\frac{\ddot{\phi}}{H\dot{\phi}}\right|\ll1,\quad \left|\frac{\dot{F}(\phi)}{HF{(\phi)}}\right|\ll1,\quad \left|\frac{\dot{G}(\phi)}{HG(\phi)}\right|\ll1,
\ea
it is not difficult to check that the effective equations obtained from action (\ref{effaction3}) agree with Eqs. (\ref{lelFreq}), (\ref{lelKGeq}) and (\ref{lelRaeq}). Therefore, in the sector with $\cos b\rightarrow-1$, the coupling functions appearing in action (\ref{effaction3}) can be regarded as the quantum effective version of their classical counterparts in the slow-roll limit.
\section{Effective dynamics of BD theory}
In this section, we choose BD theory as a test field to more clearly display the  characteristics of holonomy corrections in STT. In the literature, the action of BD theory without potentials is usually given by
\ba
S=\frac{1}{2\sqrt{\kappa}}\int_{\Sigma}d^4x\sqrt{|\det(g)|}\bigg[\varphi R
-\frac{\omega}{\varphi}(\partial^{\mu}\varphi)\partial_{\mu}\varphi\bigg],\label{BDaction}
\ea
where $\omega$ is a free dimensionless coupling parameter. For simplicity, in this section we only consider the $\omega>0$ case. Following \cite{Fujii:2003}, we call the theory of action (\ref{BDaction}) the ``prototype of BD theory," where the adjective ``prototype" emphasizes the originality of (\ref{BDaction}) compared with its many other extended versions. In the prototype of BD theory, $F(\varphi)=\sqrt{\kappa}\varphi$ and $K(\varphi)=\frac{\omega}{\sqrt{\kappa}\varphi}$; thus, we have
\ba
\frac{K(\varphi)}{G(\varphi)}=\frac{2\omega}{\sqrt{\kappa}(3+2\omega)}\frac{1}{\varphi}.\label{BDKG}
\ea
  However, the action of the operator $\widehat{\big(\frac{1}{\varphi}\big)}$ is ill defined at $\varphi=0$ in the quantization prescription we choose for the scalar field. Hence, the effective Hamiltonian (\ref{effH}) does not apply. To avoid this trouble, we introduce a new field $\phi$ by putting
  \ba
  \sqrt{\kappa}\varphi:=e^{\sqrt{\kappa}\phi};\label{fredef}
  \ea
  thus, the action of BD theory becomes
  \ba
  S^{(BD)}=
  \frac{1}{2\kappa}\int_{\Sigma}d^4x\sqrt{|\det(g)|}\bigg[e^{\sqrt{\kappa}\phi}R
-\omega e^{\sqrt{\kappa}\phi}(\partial^{\mu}\phi)\partial_{\mu}\phi\bigg].
\label{BDaction2}
  \ea
Accordingly, we get
\ba
\frac{K(\phi)}{G(\phi)}=\frac{2\omega}{3+2\omega}e^{-\sqrt{\kappa}\phi},\quad
\frac{F(\phi)}{G(\phi)}=\frac{2}{3+2\omega}e^{-\sqrt{\kappa}\phi},\quad
\frac{F'(\phi)}{G(\phi)}=\frac{2\sqrt{\kappa}}{3+2\omega}e^{-\sqrt{\kappa}\phi},\ea
of which the corresponding operators are well defined for all $\phi$. In this way, the results in the last section can be applied to BD theory.

Since $e^{\sqrt{\kappa}\phi}$ does not vanish for any $\phi$, the Friedmann equation can also be written in the form of (\ref{qFr2}) which in BD theory becomes
\ba
\bigg(e^{\sqrt{\kappa}\phi}H+\frac{1}{2}\dot{(e^{\sqrt{\kappa}\phi})}\cos b\bigg)^2
=\frac{\kappa}{3}\rho_e\bigg(1-\frac{\rho_e}{\rho_c}\bigg),\label{qFrBD}
\ea
where
\ba
\rho_e:=\frac{3+2\omega}{4}\Big(e^{\sqrt{\kappa}\phi}\dot{\phi}\Big)^2.
\label{rhoeBD}
\ea
 The Klein-Gordon equation in BD theory can be directly obtained from (\ref{qKGeq}),
 \ba
 \ddot{\phi}+3H\dot{\phi}+\frac{\dot{(e^{\sqrt{\kappa}\phi}})}{e^{\sqrt{\kappa}\phi}}\dot{\phi}
 =0,\label{qKGeqBD}
 \ea
 which translates into
 \ba
 \frac{d}{dt}\Big(|v|e^{\sqrt{\kappa}\phi}\dot{\phi}\Big)=0,\label{vedotphi}
 \ea
 from which we get $|v|e^{\sqrt{\kappa}\phi}\dot{\phi}=C$ where $C$ is a constant. Therefore, if $\dot{\phi}>0$ at the initial time, $\dot{\phi}$ will be greater than zero during the whole evolution and vice versa; hence, the theory can be divided into two independent sectors by $\dot{\phi}>0$ and $\dot{\phi}<0$. Since in the first sector $\phi$ will increase monotonically with respect to the proper time, $\phi$ can be regarded as a global internal time variable in this sector. For no matter which sector, from (\ref{qFrBD}) and (\ref{cos2b}), we infer that the Hubble parameter can vanish only at $\cos b=0$.

 Besides, Eq. (\ref{qKGeqBD}) can also be expressed as
  \ba
  \frac{d}{dt}\Big(e^{\sqrt{\kappa}\phi}\dot{\phi}\Big)=-3He^{\sqrt{\kappa}\phi}\dot{\phi}.
  \label{dedotphi}
  \ea
Plugging (\ref{rhoeBD}) into (\ref{dedotphi}), we have
\ba
\dot{\rho_e}=-6H\rho_e,
\ea
which shows that the sign of time variation of the energy density is opposite of the sign of Hubble parameter. Hence, in BD theory, the effective energy density of the scalar field in a contracting Universe will keep increasing until $\rho_e=\rho_c$; and the effective energy density in an expanding Universe will keep decreasing. Since the Hubble parameter can only vanish at $\rho_e=\rho_c$, an expanding Universe will never undergo a recollapse in BD theory.

Moreover, using (\ref{eomb}), we get
\ba
\dot{b}&=&-\frac{2\text{sgn}(v)}{(3+2\omega)\hbar}
\Bigg[\frac{\sqrt{3\Delta}}{\kappa\gamma^2}\omega
\sin^2 b
+\frac{2\sqrt{3}}{(\Delta)^{\frac{3}{2}}}\frac{\pi^2}{|v|^2}\Bigg]e^{-\sqrt{\kappa}\phi}
,\label{eombBD}
\ea
from which it is clear that $\dot{\cos b}\neq0$ at $\cos b=0$; thus, according to the arguments in Sec. III, both the sign of $\cos b$ and the sign of $H$ are bound to change around $\cos b=0$. Hence, in BD theory, the contracting branch of the Universe is always connected with the expanding branch of the Universe by the quantum bounce.

From Eq. (\ref{qFrBD}), we find
\ba
e^{\sqrt{\kappa}\phi}H+\frac{\sqrt{\kappa}}{2}e^{\sqrt{\kappa}\phi}\dot{\phi}\cos b=
\pm\sqrt{\frac{\kappa}{3}\rho_e\bigg(1-\frac{\rho_e}{\rho_c}\bigg)}.\label{lqFrBD}
\ea
Plugging Eq. (\ref{rhoeBD}) into Eq. (\ref{lqFrBD}) and using
\ba
\cos b=\text{sgn}(\cos b)\sqrt{1-\frac{\rho_e}{\rho_c}},
\ea
we get
\ba
H=\frac{\sqrt{\kappa}}{2}
\Bigg(\pm\sqrt{\frac{3+2\omega}{3}}-\text{sgn}(\cos b)\Bigg)\dot{\phi}\sqrt{1-\frac{\rho_e}{\rho_c}}
.\label{lH}
\ea
Since $\dot{\phi}$ is either positive or negative during the entire evolution and the sign of $\cos b$ is either $1$ or $-1$ in a certain branch of the Universe, the evolution of the Universe in a certain branch can be classified into four cases by the sign of $\dot{\phi}$ together with the sign of $\cos b$.

First, we consider the two cases with $\dot{\phi}>0$. In these cases, using Eq. (\ref{dedotphi}), we find that in the expanding branch of the Universe Eq. (\ref{lH}) becomes
\ba
-\frac{1}{3}\frac{d}{dt}\Big(e^{\sqrt{\kappa}\phi}\dot{\phi}\Big)
=\frac{\sqrt{\kappa}}{2}\Bigg(\sqrt{\frac{3+2\omega}{3}}
-\text{sgn}(\cos b)\Big|_{H>0}\Bigg)e^{\sqrt{\kappa}\phi}\dot{\phi}^2
\sqrt{1-\frac{3+2\omega}{4\rho_c}\Big(e^{\sqrt{\kappa}\phi}\dot{\phi}\Big)^2},
\label{expH}\ea
while in the contracting branch of the Universe Eq. (\ref{lH}) becomes
\ba
-\frac{1}{3}\frac{d}{dt}\Big(e^{\sqrt{\kappa}\phi}\dot{\phi}\Big)
&=&-\frac{\sqrt{\kappa}}{2}\Bigg(\sqrt{\frac{3+2\omega}{3}}
+\text{sgn}(\cos b)\Big|_{H<0}\Bigg)e^{\sqrt{\kappa}\phi}\dot{\phi}^2
\sqrt{1-\frac{3+2\omega}{4\rho_c}\Big(e^{\sqrt{\kappa}\phi}\dot{\phi}\Big)^2}\nn\\
&=&-\frac{\sqrt{\kappa}}{2}\Bigg(\sqrt{\frac{3+2\omega}{3}}
-\text{sgn}(\cos b)\Big|_{H>0}\Bigg)e^{\sqrt{\kappa}\phi}\dot{\phi}^2
\sqrt{1-\frac{3+2\omega}{4\rho_c}\Big(e^{\sqrt{\kappa}\phi}\dot{\phi}\Big)^2},
\label{contH}\ea
where in the last step of Eq. (\ref{contH}) we have used the fact that $\text{sgn}(\cos b)\Big|_{H<0}=-\text{sgn}(\cos b)\Big|_{H>0}$.

Due to the complexities of Eqs. (\ref{expH}) and (\ref{contH}), the analytical solution of $H(t)$ does not generally exist. However, considering that $\phi$ can be treated as an internal time variable, we may as well express the solution in terms of $\phi$. To this aim, we denote
 \ba
 f(\phi)= e^{\sqrt{\kappa}\phi}\dot{\phi};\label{fphi}
 \ea
then, from Eq. (\ref{dedotphi}) we find the Hubble parameter can be expressed by
\ba
H(\phi)=-\frac{1}{3}
\bigg(\frac{d f(\phi)}{d\phi}\bigg)e^{-\sqrt{\kappa}\phi},\label{Hf}
\ea
and thus Eqs. (\ref{expH}) and (\ref{contH}) can be rewritten as
\ba
-\frac{1}{3}\frac{d f(\phi)}{d\phi}
=\pm\frac{\sqrt{\kappa}}{2}\Bigg(\sqrt{\frac{3+2\omega}{3}}
-\text{sgn}(\cos b)\Big|_{H>0}\Bigg)f(\phi)
\sqrt{1-\frac{3+2\omega}{4\rho_c}f^2(\phi)},\label{dfdphi}
\ea
 in which the sign $``+"$ in front of the parenthesis corresponds to the expanding branch and the sign $``-"$ corresponds to the contracting branch. It is not difficult to solve Eq. (\ref{dfdphi}). For the $\text{sgn}(\cos b)\big|_{H>0}=1$ case, the solution of Eq. (\ref{dfdphi}) describing both the contracting branch and the expanding branch of the Universe is given by
\ba
f_{1}(\phi)=4\sqrt{\frac{\rho_c}{3+2\omega}}
\frac{\exp\Big[\frac{c+3}{2}\sqrt{\kappa}(\phi-\phi_b)\Big]}
{\exp\big[c\sqrt{\kappa}(\phi-\phi_b)\big]+\exp\big[3\sqrt{\kappa}(\phi-\phi_b)\big]},
\label{f1phi}
\ea
where $c\equiv\sqrt{9+6\omega}$ and $\phi_b$ denotes the value of $\phi$ at the instant of bounce.

For the $\text{sgn}(\cos b)\big|_{H>0}=-1$ case, the solution is given as follows:
\ba
f_{2}(\phi)=4\sqrt{\frac{\rho_c}{3+2\omega}}
\frac{\exp\Big[\frac{c-3}{2}\sqrt{\kappa}(\phi-\phi_b)\Big]}
{\exp\big[c\sqrt{\kappa}(\phi-\phi_b)\big]+\exp\big[-3\sqrt{\kappa}(\phi-\phi_b)\big]}.
\label{f2phi}\ea

\begin{figure}[h]
\centering
\includegraphics[height=4.8cm]{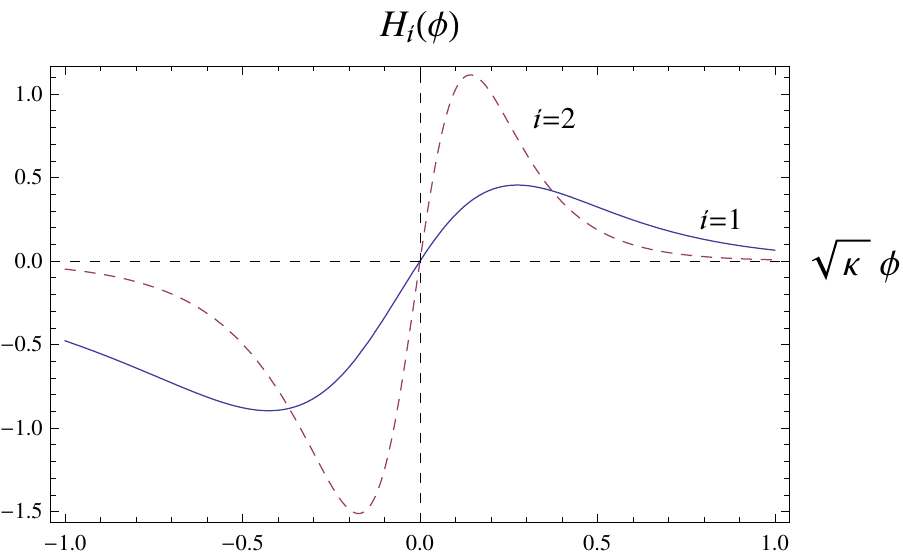}
\hspace{0.1cm}
\includegraphics[height=4.8cm]{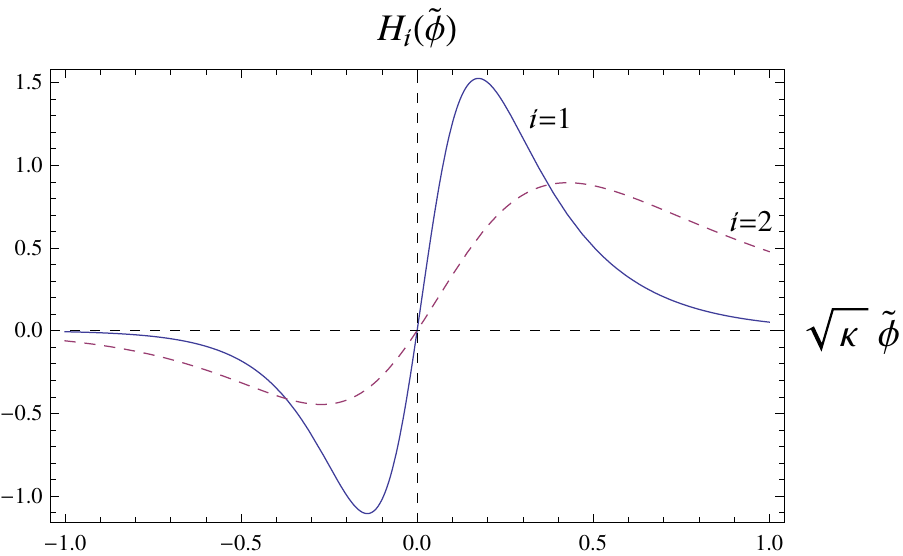}
\caption{Evolution of the Hubble parameter (in Planck units) with respect to the internal time around the bounce in different cases. For clearer comparison, in both panels, we set $\omega=10$ and $\phi_b=\tilde{\phi}_b=0$.}
\label{fig1}
\end{figure}

Using Eq. (\ref{Hf}), for $\text{sgn}(\cos b)\big|_{H>0}=1$, we obtain the solution of Hubble parameter in terms of $\phi$,
\ba
H_{1}(\phi)=\frac{2}{3}(c-3)\sqrt{\frac{\kappa\rho_c}{3+2\omega}}
\frac{\exp\Big[\frac{3c+1}{2}\sqrt{\kappa}(\phi-\phi_b)\Big]-\exp\Big[\frac{c+7}{2}\sqrt{\kappa}(\phi-\phi_b)\Big]}
{\exp(\sqrt{\kappa}\phi_b)\Big(\exp\big[c\sqrt{\kappa}(\phi-\phi_b)\big]+\exp\big[3\sqrt{\kappa}(\phi-\phi_b)\big]\Big)^2},
\label{H1phi}
\ea
while for $\text{sgn}(\cos b)\big|_{H>0}=-1$ we obtain
\ba
H_{2}(\phi)=\frac{2}{3}(c+3)\sqrt{\frac{\kappa\rho_c}{3+2\omega}}
\frac{\exp\Big[\frac{3c-5}{2}\sqrt{\kappa}(\phi-\phi_b)\Big]-\exp\Big[\frac{c-11}{2}\sqrt{\kappa}(\phi-\phi_b)\Big]}
{\exp(\sqrt{\kappa}\phi_b)\Big(\exp\big[c\sqrt{\kappa}(\phi-\phi_b)\big]+\exp\big[-3\sqrt{\kappa}(\phi-\phi_b)\big]\Big)^2}.
\label{H2phi}
\ea
Comparison of the evolution of Hubble parameter around the bounce is illustrated in the left panel of Fig. \ref{fig1}.

In the other two cases with $\dot{\phi}<0$, we denote $\tilde{\phi}\equiv-\phi$. Since $\tilde{\phi}$ monotonically increases in these cases, it can also be treated as a global time variable.
Denoting
 \ba
 g(\tilde{\phi})= e^{-\sqrt{\kappa}\tilde{\phi}}\dot{\tilde{\phi}};\label{gphi}
 \ea
 we have
 \ba
 H(\tilde{\phi})=
 -\frac{1}{3}
\bigg(\frac{d g(\tilde{\phi})}{d\tilde{\phi}}\bigg)e^{\sqrt{\kappa}\tilde{\phi}}.
\label{Hf2}
 \ea
 Then, by simply repeating the procedures above, we can directly obtain the solutions. For the $\text{sgn}(\cos b)\big|_{H>0}=1$ case, we have
\ba
g_{1}(\tilde{\phi})=4\sqrt{\frac{\rho_c}{3+2\omega}}
\frac{\exp\Big[\frac{c-3}{2}\sqrt{\kappa}(\tilde{\phi}-\tilde{\phi}_b)\Big]}
{\exp\Big[c\sqrt{\kappa}(\tilde{\phi}-\tilde{\phi}_b)\Big]+\exp\Big[-3\sqrt{\kappa}(\tilde{\phi}-\tilde{\phi}_b)\Big]},
\label{g1tildephi}
\ea
where $\tilde{\phi}_b$ denotes the value of $\tilde{\phi}$ at the bounce; and for the $\text{sgn}(\cos b)\big|_{H>0}=-1$ case we have
\ba
g_{2}(\tilde{\phi})=4\sqrt{\frac{\rho_c}{3+2\omega}}
\frac{\exp\Big[\frac{c+3}{2}\sqrt{\kappa}(\tilde{\phi}-\tilde{\phi}_b)\Big]}
{\exp\Big[c\sqrt{\kappa}(\tilde{\phi}-\tilde{\phi}_b)\Big]+\exp\Big[3\sqrt{\kappa}(\tilde{\phi}-\tilde{\phi}_b)\Big]}.
\label{g2tildephi}
\ea
The solution of Hubble parameter in each case is given by $H_{1}(\tilde{\phi})$ and $H_{2}(\tilde{\phi})$ respectively, which read
\ba
H_{1}(\tilde{\phi})=\frac{2}{3}(c+3)\sqrt{\frac{\kappa\rho_c}{3+2\omega}}
\frac{\exp\Big[\frac{3c-1}{2}\sqrt{\kappa}(\tilde{\phi}-\tilde{\phi}_b)\Big]
-\exp\Big[\frac{c-7}{2}\sqrt{\kappa}(\tilde{\phi}-\tilde{\phi}_b)\Big]}
{\exp\Big(-\sqrt{\kappa}\tilde{\phi}_b\Big)
\Big(\exp\Big[c\sqrt{\kappa}(\tilde{\phi}-\tilde{\phi}_b)\Big]+\exp\Big[-3\sqrt{\kappa}(\tilde{\phi}-\tilde{\phi}_b)\Big]\Big)^2},
\label{H1tildephi}
\ea
and
\ba
H_{2}(\tilde{\phi})=\frac{2}{3}(c-3)\sqrt{\frac{\kappa\rho_c}{3+2\omega}}
\frac{\exp\Big[\frac{3c+5}{2}\sqrt{\kappa}(\tilde{\phi}-\tilde{\phi}_b)\Big]
-\exp\Big[\frac{c+11}{2}\sqrt{\kappa}(\tilde{\phi}-\tilde{\phi}_b)\Big]}
{\exp\Big(-\sqrt{\kappa}\tilde{\phi}_b\Big)
\Big(\exp\Big[c\sqrt{\kappa}(\tilde{\phi}-\tilde{\phi}_b)\Big]+\exp\Big[3\sqrt{\kappa}(\tilde{\phi}-\tilde{\phi}_b)\Big]\Big)^2}.
\label{H2tildephi}
\ea
See the right panel of Fig. \ref{fig1} for comparison of $H_{1}(\tilde{\phi})$ and $H_{2}(\tilde{\phi})$ around the bounce.

  In the first case with $\text{sgn}(\cos b)\big|_{H>0}=1$ and $\dot{\phi}>0$, using the field redefinition in (\ref{fredef}), we find that Eq. (\ref{H1phi}) can reproduce Eq. (5.15) in Ref. \cite{Zhang:2013}. Nevertheless, the other three cases  were not considered in Ref. \cite{Zhang:2013}.

  By comparing (\ref{H1phi}) with (\ref{H2phi}), and (\ref{H1tildephi}) with (\ref{H2tildephi}), we find that the solutions agree with each other in the large $c$ limit. But it does not necessarily mean that the choice of sign of $\cos b$ does not make much difference when $c$ becomes large in BD theory, because we have not included the potential $V(\phi)$ yet. Actually, from Eq. (\ref{qKGeq}), it is clear that different signs of $\cos b$ could lead to important differences if $F'(\phi)V(\phi)$  dominates $F(\phi)V'(\phi)$.

Although it is generally not possible to find the exact solutions in terms of the proper time, in the late epoch of the expanding branch of the Universe when $\phi$ becomes large enough, it is still possible to find some approximate solutions of the scale factor $a(t)$ and the scalar field $\phi(t)$ using the above results. In the remaining part of this section, we will derive these solutions.

First, we consider the expanding branch with both $\text{sgn}(\cos b)\big|_{H>0}=1$ and $\dot{\phi}>0$. In this case, substituting (\ref{fphi}) into Eq. (\ref{f1phi}), we obtain
\ba
\frac{d\phi(t)}{dt}=4\sqrt{\frac{\rho_c}{3+2\omega}}\frac{\exp[-\sqrt{\kappa}\phi_b]}
{\exp\Big[\frac{c-1}{2}\sqrt{\kappa}\big(\phi(t)-\phi_b\big)\Big]+\exp\Big[\frac{5-c}{2}\sqrt{\kappa}\big(\phi(t)-\phi_b\big)\Big]},
\label{dphidt}
\ea
which yields
\ba
t-t_b=\alpha\exp\bigg[\frac{c-1}{2}\sqrt{\kappa}\big(\phi(t)-\phi_b\big)\bigg]
+\beta\exp\bigg[\frac{5-c}{2}\sqrt{\kappa}\big(\phi(t)-\phi_b\big)\bigg]-(\alpha+\beta),\label{tofphi}
\ea
where $t_b$ denotes the proper time at the instant of bounce and the coefficients $\alpha$, $\beta$ read
\ba
\alpha\equiv
\frac{\sqrt{3+2\omega}\exp[\sqrt{\kappa}\phi_b]}{2\sqrt{\kappa\rho_c}(c-1)},
\quad
\beta\equiv
\frac{\sqrt{3+2\omega}\exp[\sqrt{\kappa}\phi_b]}{2\sqrt{\kappa\rho_c}(5-c)}.
\ea
For brevity, we set $t_b=0$ in the remaining part of this section. Since $c>3$, the first term on the right-hand side of Eq. (\ref{tofphi}) will dominate over the other two terms when $\phi$ becomes sufficiently large, which yields
\ba
t\simeq \alpha\exp\bigg[\frac{c-1}{2}\sqrt{\kappa}\big(\phi(t)-\phi_b\big)\bigg],\label{tappro}
\ea
from which we derive
\ba
\exp \big[\sqrt{\kappa}\phi(t)\big]\simeq\exp \big[\sqrt{\kappa}\phi_0\big]\bigg(\frac{t}{t_0}\bigg)^{\frac{2}{c-1}},\label{expphi}
\ea
where $t_0$ denotes the proper time of the current Universe and $\phi_0\equiv\phi(t)|_{t=t_0}$.

In the large $\phi$ regime, Eq. (\ref{H1phi}) can also be approximated by
\ba
H_{1}(\phi)\simeq\frac{2}{3}(c-3)\sqrt{\frac{\kappa\rho_c}{3+2\omega}}\exp\big(-\sqrt{\kappa}\phi_b\big)
\exp\bigg[-\frac{c-1}{2}\sqrt{\kappa}(\phi-\phi_b)\bigg],
\label{H1phiappro}
\ea
substituting (\ref{tappro}) into Eq. (\ref{H1phiappro}), we get
\ba
H_{1}(t)\simeq \frac{c-3}{3(c-1)}\frac{1}{t},
\ea
which directly gives
\ba
a(t)\simeq a_0\bigg(\frac{t}{t_0}\bigg)^{\frac{c-3}{3(c-1)}},\label{aoft}
\ea
where $a_0\equiv a(t)|_{t=t_0}$.

Next, let us examine the $\dot{\phi}<0$ sector with $\text{sgn}(\cos b)\big|_{H>0}=1$ . It is easy to see that Eq. (\ref{g1tildephi}) and Eq. (\ref{H1tildephi}) separately reduce to the following equations for large $\tilde{\phi}$:
\ba
g_{1}(\tilde{\phi})&\simeq&4\sqrt{\frac{\rho_c}{3+2\omega}}
\exp\Big[-\frac{c+3}{2}\sqrt{\kappa}(\tilde{\phi}-\tilde{\phi}_b)\Big],
\label{g1appro}\\
H_{1}(\tilde{\phi})&\simeq&\frac{2}{3}(c+3)\sqrt{\frac{\kappa\rho_c}{3+2\omega}}
\exp\Big(\sqrt{\kappa}\tilde{\phi}_b\Big)\exp\Big[-\frac{c+1}{2}\sqrt{\kappa}(\tilde{\phi}-\tilde{\phi}_b)\Big].
\label{H1appro}
\ea
Substituting Eq. (\ref{gphi}) into Eqs. (\ref{g1appro}) and (\ref{H1appro}), after direct calculation, we obtain
\ba
\exp \big[\sqrt{\kappa}\tilde{\phi}(t)\big]&\simeq&\exp \big[\sqrt{\kappa}\tilde{\phi}_0\big]\bigg(\frac{t}{t_0}\bigg)^{\frac{2}{c+1}},\label{exptildephi}\\
a(t)&\simeq& a_0\bigg(\frac{t}{t_0}\bigg)^{\frac{c+3}{3(c+1)}},\label {aoft2}
\ea
where $\tilde{\phi}_0\equiv\tilde{\phi}(t)|_{t=t_0}$; then, using $\tilde{\phi}\equiv-\phi$, Eq. (\ref{exptildephi}) can be rewritten as
\ba
\exp \big[\sqrt{\kappa}\phi(t)\big]&\simeq&\exp \big[\sqrt{\kappa}\phi_0\big]\bigg(\frac{t}{t_0}\bigg)^{-\frac{2}{c+1}}.\label{expphi2}
\ea

Finally, by simply following the procedures above, we obtain the solutions in the cases with $\text{sgn}(\cos b)\big|_{H>0}=-1$, which read
\ba
\exp \big[\sqrt{\kappa}\phi(t)\big]&\simeq&\exp \big[\sqrt{\kappa}\phi_0\big]\bigg(\frac{t}{t_0}\bigg)^{s_{\pm}},\label{expphi3}\\
a(t)&\simeq& a_0\bigg(\frac{t}{t_0}\bigg)^{q_{\pm}},\label{aoft3}
\ea
where
\ba
s_{\pm}=\frac{2}{5\pm c},\quad q_{\pm}=\frac{1}{3}\frac{c\pm3}{c\pm5}.\label{sppm}
\ea
The solution corresponding to $(s_{+}, q_{+})$ is associated with increasing $\phi$ and the solution corresponding to $(s_{-}, q_{-})$ is associated with decreasing $\phi$. It should be pointed out that the latter solution is true only for $c>5$; when $c\in(3,5)$, derivation shows that the Hubble parameter will go to infinity in a very short time, which apparently cannot describe the evolution of our Universe.

Remarkably, in the cases with $\text{sgn}(\cos b)\big|_{H>0}=1$, both the solutions (\ref{expphi}), (\ref{aoft}) associated with increasing $\phi$ and the solutions (\ref{aoft2}), (\ref{expphi2}) associated with decreasing $\phi$ exactly agree with the O'Hanlon and Tupper solutions derived in \cite{O'Hanlon:1972}.  Note that our solutions are obtained using large $\phi$ (or $\tilde{\phi}$) approximation, while such approximation was not necessary in \cite{O'Hanlon:1972}, which seems contradictory, but actually, not. Let us offer an explanation: From (\ref{dphidt}) we learn that $\dot{\phi}$ becomes sufficiently small when $\phi$ becomes sufficiently large; thus, the effective energy density of the scalar field becomes negligible compared with $\rho_c$, and $|\cos b|\rightarrow1$. Then, if the sign of $\cos b$ is positive in the expanding branch, the effective Friedmann equation can reduce to the classical Friedmann equation from which the O'Hanlon and Tupper solutions can be obtained; however, if the sign of $\cos b$ is negative in the expanding branch, the classical Friedmann equation cannot be recovered, and this is why we get different solutions in (\ref{expphi3}) and (\ref{aoft3}).

\section{Conclusion and remarks}
In this paper, we performed a preliminary investigation of the cosmological effective dynamics with holonomy corrections in loop quantum scalar tensor theory in the Jordan frame. Now, we summarize what has been achieved. In Sec. II, the connection dynamics in terms of Ashtekar variables was developed, which is then used to get the classical evolution equations on the spatially flat FRW background. In Sec. III, the connection dynamics was quantized following the holonomy quantization prescription in standard LQC; then, the effective Hamiltonian was obtained in the timeless path integral framework, from which the effective Friedmann equation, the effective Klein-Gordon equation and the  effective Raychaudhuri equation are derived. From these equations, we found that there exists a set of evolutions equations different from the classical equations in the low-energy limit with $\text{sgn}(\cos b)=-1$, which represents another important effect of holonomy corrections in STT in addition to the contributions from the higher powers of extrinsic curvature. In Sec. IV, the BD theory without potentials was chosen as an example to concretely show the features of holonomy corrections in STT. It is found that the evolution of Universe in BD theory can be classified into four different cases by the sign of $\dot{\phi}$ and the sign of $\cos b$. In the four cases, we express the Hubble parameter as a function of the internal time variable. The differences between the evolution of the Hubble parameter around the bounce in these cases are illustrated in Fig. \ref{fig1}. Then, we obtained the solution of the scaler field and the solution of the scale factor in terms of the proper time in the large $\phi$ (or $\tilde{\phi}$) regime. In the cases with $\text{sgn}(\cos b)\big|_{H>0}=1$, the solutions coincide with the O'Hanlon and Tupper solutions, while in the cases with $\text{sgn}(\cos b)\big|_{H>0}=-1$ the solutions do not.
Finally, let us make some remarks at the end of this paper.

First, the existence of two different sets of evolution equations in the low-energy limit of STT is one of the main results of this paper. Since it is widely believed that there exists a slow-roll inflation which takes place at a energy density much lower than the scale of Planck energy density, an interesting question arises: Can the set of equations of motion (\ref{lelFreq}), (\ref{lelKGeq}) and (\ref{lelRaeq}) associated with $\cos b\rightarrow-1$  also describe the evolution of our Universe after the beginning of slow-roll inflation in some specific models of STT? And, what are the differences between the physical predictions of the two sectors? To answer these questions, much more work needs to be done. For instance, we can check whether or not the slow-roll inflation can take place in some widely studied specific models of STT in both sectors. Since the quantum bounce does not necessarily exist in LQC of STT, to study the slow-roll inflation in these specific models, following \cite{Linsefors2013}, we have to set the initial conditions in the remote past in a contracting Universe and then check if a flat probability distribution function can be assigned during the contracting phase of the Universe. If the slow-roll inflation with e-folds number $N>60$ can probably take place in both sectors in some specific models, then we can use the cosmological perturbation theory to calculate the spectral indices in both sectors and check whether they both lie in the observation range. To this aim, first we can develop the cosmological perturbation theory in LQC of STT following the commonly used approaches such as the dressed metric approach \cite{Agullo2012} or the deformed algebra approach (see, for instance, the Refs. \cite{Bojowald2008,Cailleteau2012,Han2018}). These works will be our main concern in the future research.

Moreover, it is also worth investigating that whether the two sectors separately associated with $\cos b\rightarrow1$ and $\cos b\rightarrow-1$ are dynamically independent or not in some specific models of STT. To clarify this issue, we can set the initial conditions in the low-energy limit of one sector in a contracting Universe; then, we can study the evolution of $\cos b$ and see whether it can evolve into the low-energy limit of the other sector. Obviously, the necessary condition to realize the evolution from one sector to the other is that $\cos b$ must reach zero and then cross it, i.e. the quantum bounce must take place during the evolution. Roughly speaking, the evolution of $\cos b$ can be classified into three cases. First, for those particular models in which bouncing behaviors take place at a maximum energy density much lower than $\rho_c$ (such as the models considered in Refs. \cite{Boisseau:2015,Pozdeeva:2016}), the quantum bounce can never happen and the evolution of $\cos b$ always remains in the sector with $\cos b\rightarrow1$, nonetheless, it is still a question whether the evolution of $\cos b$ can remain in the other sector with $\cos b\rightarrow-1$ in these models. Second, for those models in which neither the above classical-like bounce nor the quantum bounce could take place (such as the model considered in Ref. \cite{Chen:2018}), the solutions will flow to some fixed points in the contracting phase of the Universe, in this case, the evolution of $\cos b$ also remains in a certain sector. Third, for those models in which the quantum bounce can take place (such as the BD theory considered in our paper), the sign of $\cos b$ changes during the evolution, in this case, we need to check whether the solutions of these models can stably flow to the low-energy limit of the other sector or to some other fixed points in an expanding Universe. To summarize, whether the two sectors are related to each other depends on the specific model we choose; and it is only after analyzing the equations of motion obtained in this paper that we can determine to which case the evolution of $\cos b$ in a specific model belongs.

Second, we would like to discuss more about the quantization prescription used in this paper. For the quantization of connection, we use the prescription proposed in \cite {Ashtekar:2009}. In recent years, another proposal of quantization of Hamiltonian constraint formulated in \cite{Yang:2009,Dapor:2018,Assanioussi:2018} resembling the holonomy quantization in the full theory arouses much interest among LQC community. It is worth investigating which new effects the application of this new quantization scheme would bring about in STT, especially in the low-energy limit; and this will be left for future research. For the quantization of the scalar field, we choose the Schr\"{o}dinger representation, which requires that the functions $\frac{K(\phi)}{G(\phi)}$, $\frac{F'(\phi)}{G(\phi)}$, $\frac{F(\phi)}{G(\phi)}$, $V(\phi)$ should have no singularities for any $\phi$. However, for some commonly used coupling functions, this requirement cannot be satisfied. Although this problem can sometimes be avoided by introducing a new scalar field such as in Sec. IV, we often lose the evolution information in some interval of the original scalar field. Perhaps adoption of polymer quantization for the scalar field in further study can help to alleviate this problem.

\begin{acknowledgements}
We thank Prof. Yongge Ma for illuminating conversations and Dr. Long Chen for helpful discussions. This work is supported by NSFC (No. 11905178 and No. 11475143) and Nanhu Scholars Program for Young Scholars of Xinyang Normal University.
\end{acknowledgements}


\begin{thebibliography}{99}
\bibitem{Bergmann:1968}
P. G. Bergmann, Comments on the scalar-tensor theory, Int. J. Theor. Phys. {\bf1}, 25 (1968). 

\bibitem{Steinhardt:1990}
P. J. Steinhardt and F. S. Accetta, Hyperextended Inflation, Phys. Rev. Lett. {\bf64}, 2740 (1990).

\bibitem{Garc¨ªa-Bellido:1990}
J. Garc¨ªa-Bellido and M. Quir\'{o}s, Extended inflation in scalar-tensor theories of gravity, Phys. Lett. B {\bf243}, 45 (1990).

\bibitem{Boisseau:2000}
B. Boisseau, G. Esposito-Far¨¨se, D. Polarski, and A. A. Starobinsky, Reconstruction of a Scalar-Tensor Theory of Gravity in an Accelerating Universe, Phys. Rev. Lett. {\bf85}, 2236 (2000). [arXiv:gr-qc/0001066].

\bibitem{Elizalde:2004}
E. Elizalde, S. Nojiri, and S. D. Odintsov, Late-time cosmology in a (phantom) scalar-tensor theory: Dark energy and the cosmic speed-up, Phys. Rev. D {\bf70}, 043539 (2004). [arXiv:hep-th/0405034].

\bibitem{Faraoni:2004}
V. Faraoni, \textit{Cosmology in Scalar-Tensor Gravity} (Kluwer Academic Publishers, Dordrecht, 2004)

\bibitem{Ade:2014}
P. A. R. Ade \emph{et al}., Planck 2013 results. XXII. Constraints on inflation, Astron. Astrophys. {\bf571}, A22 (2014). [arXiv:1303.5082[astro-ph.CO]].

Planck 2015 results-XX. Constraints on inflation, Astron. Astrophys. {\bf594}, A20 (2016). [arXiv:1502.02114 [astro-ph.CO]].

\bibitem{Sebastiani:2014}
L. Sebastiani, G. Cognola,  R. Myrzakulov,  S. D. Odintsov, and S. Zerbini,  Nearly Starobinsky inflation from modified gravity. Phys. Rev. D {\bf89}, 023518 (2014).	[arXiv:1311.0744 [gr-qc].

\bibitem{Aref¡¯eva:2014}
I. Y. Aref¡¯eva,  N. V. Bulatov,   R. V. Gorbachev, and S. Y. Vernov,  Non-minimally coupled cosmological models with the Higgs-like potentials and negative cosmological constant. Classical Quantum Gravity {\bf31}, 065007 (2014). [arXiv:1206.2801[gr-qc]]. 

\bibitem{Dom¨¨nech:2015}
G. Dom¨¨nech and M. Sasaki, Conformal frame dependence of inflation, J. Cosmol. Astropart. Phys. {\bf04} (2015) 022. [arXiv:1501.07699 [gr-qc]]. 

\bibitem{Ashtekar:2006}
A. Ashtekar, T. Paw\l{}owski, and P. Singh, Quantum nature of the big bang: Improved dynamics, Phys. Rev. D {\bf74}, 084003 (2006). [arXiv:gr-qc/0607039]. 

Quantum Nature of the Big Bang, Phys. Rev. Lett. {\bf96}, 141301 (2006). [arXiv:gr-qc/0602086]. 

Quantum nature of the big bang: An analytical and numerical investigation, Phys. Rev. D {\bf73}, 124038 (2006). [arXiv:gr-qc/0604013]. 

\bibitem{Ashtekar:2011}
A. Ashtekar and P. Singh,  Loop quantum cosmology: A status report, Classical Quantum Gravity {\bf 28}, 213001 (2011).[arXiv:1108.0893[gr-qc]].

\bibitem{Bojowald:2005}
M. Bojowald, Loop quantum cosmology, Living Rev. Relativity. {\bf8}, 11 (2005).[arXiv:gr-qc/0601085]. 

\bibitem{Barrau:2014}
A. Barrau, M. Bojowald, G. Calcagni, J. Grain, and M. Kagan, Anomaly-free cosmological perturbations in effective canonical quantum gravity,  J. Cosmol. Astropart. Phys. {\bf05} (2015) 051. [arXiv:1404.1018[gr-qc]].

\bibitem{Veraguth:2017}
O. J. Veraguth and C. H.-T. Wang, Immirzi parameter without Immirzi ambiguity: Conformal loop quantization of scalar-tensor gravity, Phys. Rev. D {\bf96}, 084011 (2017). [arXiv:1705.09141 [gr-qc]]. 

\bibitem{Bojowald:2006}
M. Bojowald and M. Kagan, Loop cosmological implications of a nonminimally coupled scalar field, Phys. Rev. D {\bf74}, 044033 (2006). [arXiv:gr-qc/0606082v1];

Singularities in isotropic non-minimal scalar field models, Classical Quantum Gravity {\bf23}, 4983 (2006).[arXiv:gr-qc/0604105]. 

\bibitem{Artymowski:2012}
M. Artymowski, A. Dapor, and T. Pawlowski, Inflation from nonminimally coupled scalar field in loop quantum cosmology, J. Cosmol. Astropart. Phys. {\bf06},(2013) 010.
[arXiv:1207.4353[gr-qc]].

\bibitem{Amor¨®s:2014}
J. Amor¨®s, J. de Haro, and S. D. Odintsov, On $R+\alpha R^2$ loop quantum cosmology, Phys. Rev. D {\bf89}, 104010 (2014). [arXiv:1402.3071[gr-qc]]. 

\bibitem{Odintsov:2014}
S. D. Odintsov and V. K. Oikonomou, Matter bounce loop quantum cosmology from $F(R)$ gravity, Phys. Rev. D {\bf90}, 124083 (2014). [arXiv:1410.8183 [gr-qc]].

\bibitem{Jin:2018}
W. Jin, Y. Ma, and T. Zhu, Pre-inflationary dynamics of Starobinsky inflation and its generization in loop quantum Brans-Dicke cosmology, J. Cosmol. Astropart. Phys.  {\bf02} (2019) 010. [arXiv:1808.09643 [gr-qc]].

\bibitem{Haro:2018}
J. de Haro, S. D. Odintsov, and V. K. Oikonomou, Viable inflationary evolution from loop quantum cosmology scalar-tensor theory, Phys. Rev. D {\bf97}, 084052 (2018).
[arXiv:1802.09024 [gr-qc]]. 

\bibitem{Zhang:2011a}
X. Zhang and Y. Ma, Nonperturbative loop quantization of scalar-tensor theories of gravity, Phys. Rev. D {\bf84}, 104045 (2011).[arXiv:1107.5157 [gr-qc]].

\bibitem{Zhang:2011b}
X. Zhang and Y. Ma, Extension of Loop Quantum Gravity to $f(R)$ Theories, 	Phys. Rev. Lett. {\bf106}, 171301 (2011). [arXiv:1101.1752 [gr-qc]];

Loop quantum $f(R)$ theories, Phys. Rev. D {\bf84}, 064040 (2011). [arXiv:1107.4921 [gr-qc]].

\bibitem{Zhang:2013}
X. Zhang, M. Artymowski, and Y. Ma, Loop quantum Brans-Dicke cosmology,  Phys. Rev. D {\bf87}, 084024 (2013). [arXiv:1211.4183 [gr-qc]]. 

\bibitem{Artymowski:2013}
M. Artymowski, Y. Ma, and X. Zhang, A comparison between Jordan and Einstein frames of Brans-Dicke gravity a la loop quantum cosmology, Phys. Rev. D {\bf88}, 104010  (2013). [arXiv:1309.3045 [gr-qc]].


\bibitem{Chen:2018}
L. Chen, Dynamical analysis of loop quantum $R^2$ cosmology, Phys. Rev. D {\bf99}, 064025 (2019) [arXiv:1811.08235 [gr-qc]].

\bibitem{Liddle:1992}
A. R. Liddle and D. Wands, Hyperextended inflation: Dynamics and constraints, Phys. Rev. D {\bf45}, 2665 (1992).

\bibitem{Torres:1996}
D. F. Torres and H. Vucetich, Hyperextended scalar-tensor gravity,  Phys. Rev. D {\bf54}, 7373 (1996). [arXiv:gr-qc/9610022].

\bibitem{Han:2015}
Y. Han, K. Giesel, and Y. Ma, Manifestly gauge invariant perturbations of scalar-tensor theories of gravity, Classical Quantum Gravity {\bf32}, 135006 (2015). [arXiv:1501.04947 [gr-qc]].

\bibitem{Ashtekar:2004}
A. Ashtekar, J. Lewandowski, Background independent quantum gravity: A status report,
Classical Quantum Gravity {\bf21}, R53 (2004). [arXiv:gr-qc/0404018]. 

\bibitem{Thiemann:2007}
T. Thiemann, \textit{Modern Canonical Quantum General Relativity}
(Cambridge University Press, Cambridge, England, 2007).

\bibitem{Boisseau:2015}
B. Boisseau, H. Giacomini, D. Polarski, and A. A. Starobinsky, Bouncing universes in scalar-tensor gravity models admitting negative potentials,  J. Cosmol. Astropart. Phys. {\bf07} (2015) 002. [arXiv:1504.07927 [gr-qc]];

B. Boisseau, H. Giacomini, and D. Polarski, Bouncing universes in scalar-tensor gravity around conformal invariance, J. Cosmol. Astropart. Phys. {\bf05} (2016) 048.
[arXiv:1603.06648 [gr-qc]].

\bibitem{Pozdeeva:2016}
E. O. Pozdeeva, M. A. Skugoreva, A. V. Toporensky, and S. Y. Vernov, Possible evolution of a bouncing universe in cosmological models with nonminimally coupled scalar fields,
J. Cosmol. Astropart. Phys. {\bf12},(2016) 006. [arXiv:1608.08214 [gr-qc]].

A. Y. Kamenshchik, E. O. Pozdeeva, A. Tronconi, G. Venturi, and S. Y. Vernov, Interdependence between integrable cosmological models with minimal and nonminimal coupling, Classical Quantum Gravity {\bf33}, 015004 (2016). [arXiv:1509.00590 [gr-qc]].

E. O. Pozdeeva and S. Y. Vernov, Induced gravity models with exact bounce solutions, Phys. Part. Nuclei {\bf49}, 914 (2018). [arXiv:1711.06255 [gr-qc]]. 

\bibitem{Ashtekar:2009}
A. Ashtekar and E. Wilson-Ewing, Loop quantum cosmology of Bianchi type II models,
Phys. Rev. D {\bf80}, 123532 (2009). [arXiv:0910.1278 [gr-qc]].

\bibitem{Wilson-Ewing:2010}
E. Wilson-Ewing, Loop quantum cosmology of Bianchi type IX models, Phys. Rev. D {\bf82}, 043508 (2010). [arXiv:1005.5565 [gr-qc]].

\bibitem{Corichi:2011}
A. Corichi and A. Karami, Loop quantum cosmology of $k=1$ FRW: A tale of two bounces, Phys.Rev.D {\bf84}, 044003 (2011). [arXiv:1105.3724 [gr-qc]].

\bibitem{Ashtekar:2010}
A. Ashtekar, M. Campiglia, and A. Henderson, Casting loop quantum cosmology in the spin foam paradigm, Classical Quantum Gravity {\bf27}, 135020 (2010). [arXiv:1001.5147 [gr-qc]]. 

Path integrals and the WKB approximation in loop quantum cosmology, Phys. Rev. D {\bf82}, 124043 (2010). [arXiv:1011.1024 [gr-qc]]. 

\bibitem{Fujii:2003}
Y. Fujii and K. I. Maeda, \textit{The Scalar-Tensor Theory of Gravitation}
(Cambridge University Press, Cambridge, England 2003).

\bibitem{O'Hanlon:1972}
J. O'Hanlon and B. O. J. Tupper, Vacuum-field solutions in the Brans-Dicke theory, Nuovo Cimento B {\bf7}, 305 (1972). 


\bibitem{Linsefors2013}
L. Linsefors and A. Barrau, Duration of inflation and conditions at the bounce as a prediction of effective isotropic loop quantum cosmology, Phys. Rev. D {\bf87}, 123509 (2013). [arXiv:1301.1264 [gr-qc]]. 

K. Martineau, A. Barrau, and S. Schander, Detailed investigation of the duration of inflation in loop quantum cosmology for a Bianchi-I universe with different inflaton potentials and initial conditions, Phys. Rev. D {\bf95}, 083507 (2017). [arXiv:1701.02703 [gr-qc]]. 

\bibitem{Agullo2012}
I. Agullo, A. Ashtekar, and W. Nelson, Quantum Gravity Extension of the Inflationary Scenario, Phys. Rev. Lett. {\bf 109}, 251301 (2012).
[arXiv:1209.1609 [gr-qc]];

 Extension of the quantum theory of cosmological perturbations to the Planck era, Phys. Rev. D {\bf87}, 043507 (2013).	[arXiv:1211.1354 [gr-qc]];

 The preinflationary dynamics of loop quantum cosmology:Confronting quantum gravity with observations, Classical  Quantum Gravity {\bf30}, 085014 (2013). [arXiv:1302.0254 [gr-qc]].

\bibitem{Bojowald2008}
M. Bojowald, G. M. Hossain, M. Kagan, and S. Shankaranarayanan,
  Anomaly freedom in perturbative loop quantum gravity, Phys. Rev. D {\bf 78}, 063547 (2008).  [arXiv:0806.3929 [gr-qc]].

  Gauge invariant cosmological perturbation equations with corrections from
  loop quantum gravity, Phys. Rev. D {\bf 79}, 043505 (2009).  [arXiv:0811.1572 [gr-qc]].

\bibitem{Cailleteau2012}
T. Cailleteau, J. Mielczarek, A. Barrau, and J. Grain, Anomaly-free scalar perturbations with holonomy corrections in loop quantum cosmology, Classical Quantum Gravity {\bf29}, 095010 (2012). [arXiv:1111.3535 [gr-qc];	

T. Cailleteau, A. Barrau, J. Grain, and F. Vidotto, Consistency of holonomy-corrected scalar, vector and tensor perturbations in loop quantum cosmology, Phys. Rev. D {\bf86}, 087301 (2012). [arXiv:1206.6736 [gr-qc]]; 

A. Barrau, P. Jamet, K. Martineau, and F. Moulin, Scalar spectra of primordial perturbations in loop quantum cosmology, Phys. Rev. D {\bf98}, 086003 (2018).
[arXiv:1807.06047 [gr-qc]].

\bibitem{Han2018}
Y. Han, Cosmological perturbations with inverse-volume corrections in loop quantum cosmology, Phys. Rev. D {\bf98}, 083507 (2018).[arXiv:1809.00313 [gr-qc]].

\bibitem{Yang:2009}
J. Yang, Y Ding, and Y. Ma, Alternative quantization of the Hamiltonian in loop quantum cosmology II: Including the Lorentz term, Phys. Lett. B {\bf682}, 1 (2009).  [arXiv:0904.4379 [gr-qc]]. 

\bibitem{Dapor:2018}
A. Dapor and K. Liegener, Cosmological effective Hamiltonian from full loop quantum gravity dynamics, Phys. Lett. B {\bf785}, 506 (2018). [arXiv:1706.09833 [gr-qc]].

\bibitem{Assanioussi:2018}
M. Assanioussi, A. Dapor, K Liegener, and Tomasz Paw\l{}owski, Emergent de Sitter Epoch of the Quantum Cosmos, Phys. Rev. Lett. {\bf121}, 081303 (2018). [arXiv:1801.00768 [gr-qc]]. 
\end{thebibliography}
\end{document}